  \pdfoutput=1
  \documentclass[runningheads,twoside]{llncs}

\newif\iffullversion
  \fullversiontrue

\newif\ifauthversion
  \authversiontrue 

\newif\ifdraftversion
  \draftversionfalse

\newif\ifsubmission 
  \iffullversion
    \ifauthversion
      \submissionfalse
    \else
      \submissiontrue
    \fi
  \else
    \submissionfalse
\fi

\newif\ifeprintfull 
  \iffullversion
    \ifauthversion
      \eprintfulltrue
    \else
      \eprintfullfalse
    \fi
  \else
    \eprintfullfalse
  \fi

\usepackage{cite} 

\ifeprintfull
  \usepackage[a4paper,margin=1.0in]{geometry}

  \usepackage[T1]{fontenc}

  \usepackage[tt=false]{libertine}
  \usepackage[varqu]{zi4}
  \usepackage[libertine,nosymbolsc]{newtxmath}

  \usepackage[section]{placeins} 
\fi

\ifsubmission
  \usepackage[reset]{geometry}
\else
\fi

\usepackage[hyphens]{url}
\usepackage{
  lineno,
  hyperref,
  graphicx,
  amsmath,
  IEEEtrantools
}

\begin{document}

  
  \title{
    Stupid, Evil, or Both?\\
    Understanding the Smittestopp conflict  
  }
  
  \ifsubmission
    \author{
      \vspace*{-5mm}
    }
    \institute{
      \vspace*{-5mm}
    }
  \else
    \author{
      Hans Heum\inst{1} 
    }
    \authorrunning{H.~Heum}
  
    \institute{
      Simula UiB
      \iffullversion
        \\ Merkantilen (3rd floor) \\ Thorm{\o}hlensgate 53D \\ N-5006 Bergen,
      \fi
      Norway.\\
      \email{hansh@simula.no}
    }
  \fi
  
  \maketitle
  
  \begin{abstract}
    Like many governments, the Norwegian government provided a contact tracing application to help in
      combating the COVID-19 pandemic at its outset. However, the application was widely criticized
      for enabling an unacceptable intrusion into its subjects’ lives, leading to its discontinuation
      only four months into the pandemic. In this essay, we will take a closer look at what went wrong,
      attempt to gain a deeper understanding of the passionate nature of the conflict, and how both
      sides came to view the other as being either stupid, or evil, or both.\footnote{
        The following essay was originally written
        October 2020; it has been updated for publication in 2021.
      }

  \end{abstract}
  
  \ifsubmission\setcounter{page}{1}\fi

  \section{Introduction} 
  	On March 31, 2020, the preprint \emph{Quantifying SARS-CoV-2 transmission suggests epidemic control
with digital contact tracing}~\cite{ferretti2020quantifying} appeared online, having been accepted for publication in
the prestigious journal Science.

On April 16, Norwegian authorities released their solution: Smittestopp. Developed by the
Norwegian Institute of Public Health (NIPH),\footnote{Norwegian: Folkehelseinstituttet (FHI).} in collaboration with Simula Research Laboratory, the
app had the dual purposes of providing automatic, digital contact tracing, and of gathering data to
monitor the spread of the coronavirus~\cite{ss-launch}.

On June 2, Amnesty International presented an open letter to the Norwegian Minister of Justice
and Public Security, claiming that the app was violating fundamental human rights, and urging the 
Norwegian government to immediately roll back its employment~\cite{amnesty-letter}.

On June 12, the Norwegian Data Protection Authority (NDPA)\footnote{Norwegian: Datatilsynet.} announced a temporary ban on the 
processing of any and all data gathered by the app~\cite{datatilsynet-varsel}. 

On June 15, the app was discontinued~\cite{ss-end}.

To this day, all parties appear convinced of their moral high ground,
viewing the opposing side as ``either stupid, or evil, or both''.\footnote{Quoting a source close to
the project, who for the purpose of this essay shall remain anonymous, in conversation with the
author.} On the one side stands NIPH, together with the developers of Simula Research Laboratories, 
apparently firm in their belief that the 
loss of privacy was a small price to pay for data on the spreading of the
disease~\cite{kyrres-uttalelser}. On the other side stands NDPA and Amnesty International, firm in their belief that such an intrusion
would violate the Universal Declaration of Human Rights.\footnote{Specifically, Article 12; see~\cite{human-rights}.}

How did such a fundamental disagreement arise? Is there hope for reconsiliation? And, finally, who
are right? \emph{Should} Smittestopp's intrusive data gathering capabilities have been accepted in the
name of science and public health?


  \section{A Doctor's Oath}
  	\begin{figure}[t]
	\begin{center}
		\includegraphics[width=\textwidth]{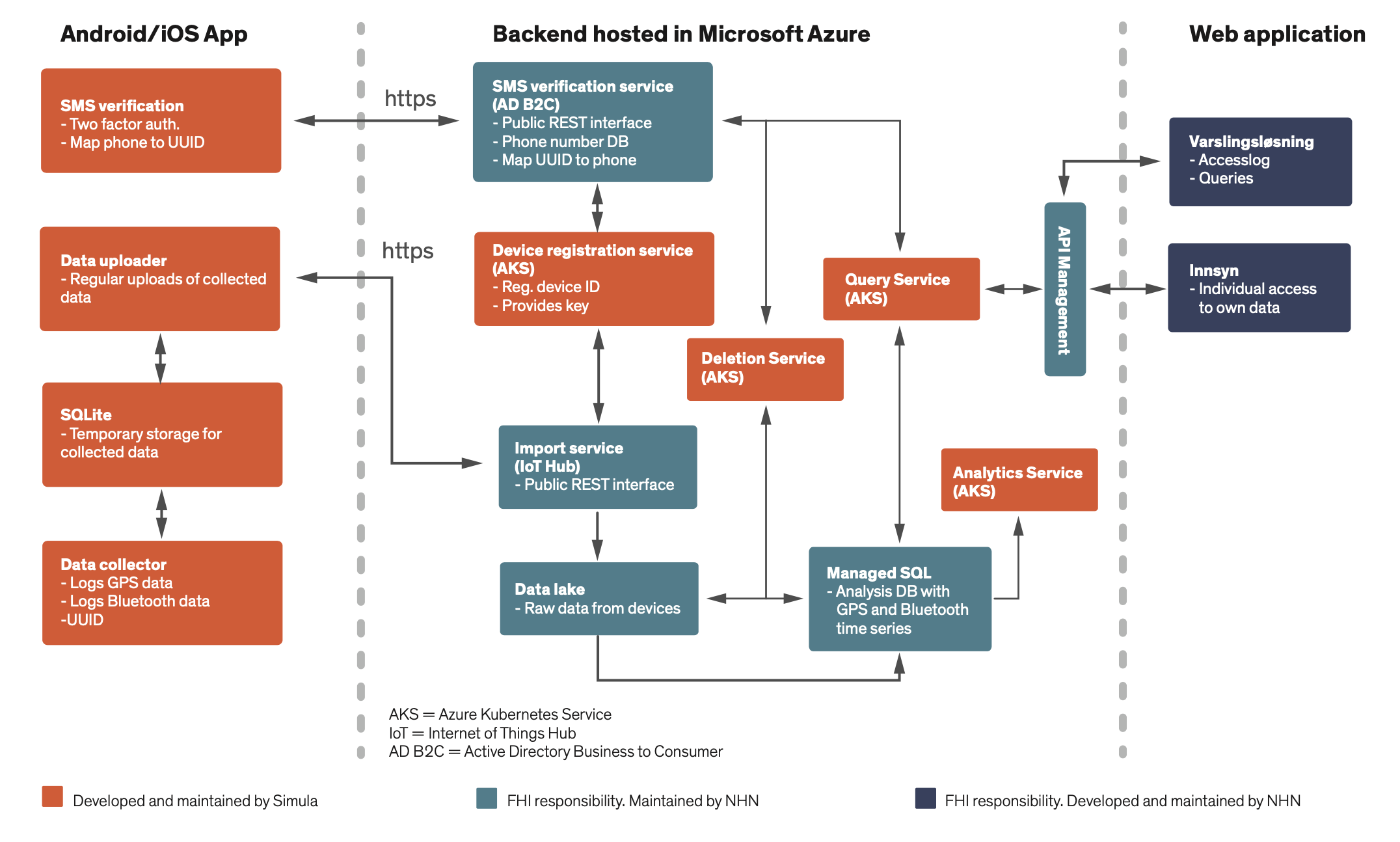}
		\caption{
			An overview of the application architecture of Smittestopp. As can be seen (bottom-left), the app collects GPS data
			and Bluetooth data, and connects them with a unique identifier (UUID). Additionally, it collects
			the users phone number, as needed for the two-factor authentication, and associates it
			with their UUID (upper-left). Note that the UUID remains constant for each user.
			It seems likely that, for someone with full access to the servers, connecting parts of the data set to specific
			users would be a trivial task.
      Figure taken from~\protect{\cite[page 11]{simula-rapport}}. The servers were located in Ireland, and
      were operated by Microsoft.\label{ss-specs}
		}
	\end{center}
\end{figure}
In attempting to understand the viewpoints of the opposing parties, it is crucial to try to
understand the respective research communities surrounding them. A useful tool towards this goal
is to look at
how \emph{codes of ethics} appear in each: this can help reveal common biases brought on
through years of training and engaging with the community---even if those same codes of
ethics are far from at the front of a researcher's mind in the middle of a heated argument.

To start, it is commonly taught that Hippocrates wrote the first binding treatise on medical ethics.
Hailing from between the 5th and the 3rd centuries BC, it included an oath to be sworn 
by all practitioners of medicine~\cite{hippocratic-oath}.
The oath spells out guidelines relating to the sharing of medical knowledge, and to the treatment and
care of patients, with the central theme being that patients should be regarded as human beings rather
than scientific subjects.

The oath survives today in modernized forms, and while several versions exist, the most widely adopted modern version of the
oath is the one written in 1964 by Dr. Louis Lasagna, Academic Dean at Tufts University School of
Medicine~\cite{modern-oath}. This is the version that will be quoted here. 

One line of the oath reads as follows:

\begin{quote}
	\emph{I will prevent disease whenever I can, for prevention is preferable to cure.}
\end{quote}

Many members of the Norwegian Institute of Public Health will likely have sworn this oath upon
graduation. If they see the potential for disease prevention, then they will quite literally have a
moral obligation to pursue it, as long as it doesn't otherwise contradict the oath, or the law at
large.

The oath is not silent on the subject of privacy either, stating:

\begin{quote}
	\emph{I will respect the privacy of my patients, for their problems are not disclosed to me that
	the world may know.}
\end{quote}

Meanwhile, in January 2020, the Nuffield Council On Bioethics released their report, ``Research in global
health emergencies: ethical issues" \cite{bioethics-rapport}. Endorsed by the World Health
Organization \cite{who-endorsement},
the report has become a guiding document to research on the pandemic.
Chapter 9 of the report deals with the handling of data. To
quote from the chapter summary (page 186): 

\begin{quote}
	\emph{Sharing data and samples between humanitarian actors, or for future research use, can play an
	important role in helping reduce suffering in many ways, both during emergencies and in the routine
	surveillance that forms part of emergency preparedness. However, sharing may also bring with it
	risks of harm and exploitation} [\dots] \emph{Sharing is vital for effective research collaboration,
	but it must not be exploitative.
	}
\end{quote}

Reading this, one is reminded of the first promise of the Hippocratic oath: 

\begin{quote}
	\emph{I will respect the hard-won scientific gains of those physicians in whose steps I walk, and
	gladly share such knowledge as is mine with those who are to follow.}
\end{quote}

In other words, they are talking about the more-or-less open sharing of potentially private or harmful data used
in research. It should be
obvious to anyone that such sharing is to be done with great care, and the main tool to guard
against this kind of misconduct, is oversight.

It was likely quite clear to the medical professionals involved with the design 
of Smittestopp that there would be no danger of such misconduct---after all, they were looking to publish 
\emph{statistics}, not the names and phone numbers of their subjects. As is common
practice, such data would be purged from the data set before analysis even began. 
Meanwhile, the data itself would remain securely hidden away on servers in Ireland, guarded
against intrusion at all hours, and the scientists involved would have careful oversight on the manners in
which the data was used in their research.

With a moral obligation to prevent the spreading of the disease, and with the collection of data
seen as vital to this effort, together with the fact that no disclosure of personal
information was to take place, one begins to understand how collecting movement data on infected
individuals was seen as an absolutely vital part of the application's purpose and design, and it's likely that the medical researchers 
didn't see the presence of an ethical dilemma here at all. On the contrary, one
imagines that they must have been genuinely confused at the backlash that was to follow.

  \section{A Human Right}
  	Article 12 of the Universal Declaration of Human Rights~\cite{human-rights} reads:

\begin{quote}
	\emph{No one shall be subjected to arbitrary interference with his privacy, family, home or
	correspondence, nor to attacks upon his honour and reputation. Everyone has the right to the
	protection of the law against such interference or attacks.}
\end{quote}
The extents and limits of this right are (as with most legal texts) subject to fervent
debate. Nevertheless, it may well be viewed as foundational for the \emph{privacy community}. 

To quote Eric Hughes' influential 1993 text, \emph{A Cypherpunk's Manifesto}~\cite{cypherpunk}:

\begin{quote}
	\emph{We cannot expect governments, corporations, or other large, faceless organizations to
grant us privacy out of their beneficence. It is to their advantage to speak of us, and we should
expect that they will speak. To try to prevent their speech is to fight against the realities of
information.} [\dots]\ \emph{We must defend our own privacy if we expect to have any.}
\end{quote}

Their greatest fear is typically an Orwellian nightmare, in which individuals are monitored and tracked,
free movement and speech are distant memories, and even free thought is under constant attack. 
Furthermore, they view the road to such a society as a slippery slope towards which all authority is inevitably
drawn.\footnote{And there is certainly historical precedent for such a view! As the saying goes, ``Power corrupts; absolute power corrupts absolutely.''} 
Therefore, one must be vigilant and nip in the bud any and all attempts by people in power
to relax privacy rights or increase surveillance efforts. Needless to say, they view all data 
collection and monitoring with extreme skepticism, regardless of its stated purpose.

On the other hand, in the modern world, it is not only state actors that infringe upon our privacy.
Modern communication technology has made it possible for companies
like Google and Facebook to build entire industries around harvesting personal data and selling
it to advertising firms. While their motivations may seem less sinister than those envisioned of
the state, the power they amass through such data harvesting\footnote{Not to mention the power gained 
through the billions earned in targeted advertising.} cannot be overstated. Such power erodes democracy itself, as can be
seen by the fact that governments now regularly plead with the leaders of big
international corporations,
as if they were sovereign entities themselves.\footnote{The subject of digital contact tracing actually provides 
a very clear example of this, as we shall see later.}

For the above reasons, many members of the privacy community will oppose \emph{any} effort to implement digital
contact tracing---even those treating the privacy of its subjects with the utmost care and
respect. On the other hand, there are many who would agree that if an application could be designed such that it is
verifiably impossible 
(or at least verifiably very difficult) for anyone, \emph{including the maintainers of
the system}, to exploit the system for anything beyond
contact tracing, \emph{then} the benefits of digital contact tracing will outweigh the tiny loss of
privacy. 

How to achieve fully privacy-preserving digital contact tracing remains an open problem, and no
widely-agreed-upon solution exists~\cite{vaudenay2020centralized}, but at a minimum, the application should have an open source code, 
and a description of all data processing should be available and clearly presented, in order that
any interested party may verify the design guarantees.\footnote{
This is usually done by trying to come up with ways to
\emph{break} them, and communicating to the maintainers any security holes they find.
Trust in the system as a whole then comes from a shared trust in its maintainers to continually roll
out security patches, and in the community to vigilantly find clever new ways to attack it.}
This, and more, is spelled out in a Joint Statement on Contact
Tracing, dated April 19, 2020, and signed by hundreds of researchers from around the world~\cite{joint-statement}.

But with Smittestopp \emph{explicitly} designed to collect and exploit data 
beyond the purpose of contact tracing, storing both personal information and movement data 
on central servers with no clear separation between the two, \emph{and} launching
with a closed source code, the app simply failed on all counts. It should not be difficult to see how the privacy
community was both shocked and enraged by its design. To the privacy enthusiast, the 
infringements represented by Smittestopp were simply unacceptable.\footnote{There are further
issues with the Smittestopp design beyond those discussed here; one is the use of static
identifiers. In a blog post, Ian Levy,
the technical director at the National Cyber Security Centre UK (NCSC), writes that ``In any contact
tracing app that broadcasts something to be picked up by others, there are risks. There are a range
of schemes from having a fixed-for-all-time ID that's constantly broadcast (which would be silly as
anyone can see if you're around), through to schemes that make it exceptionally difficult to work
out what's going on. There are a set of well known attacks that all apps have to
mitigate.''~\cite{silly}
The above paranthesis turns particularly bemusing once one realizes that this is
exactly what Smittestopp did, as can be seen in Fig.~\ref{ss-specs}.}

  \section{Late Lessons, Early Warnings}
  	\begin{figure}
	\begin{center}
		\includegraphics[width=\textwidth]{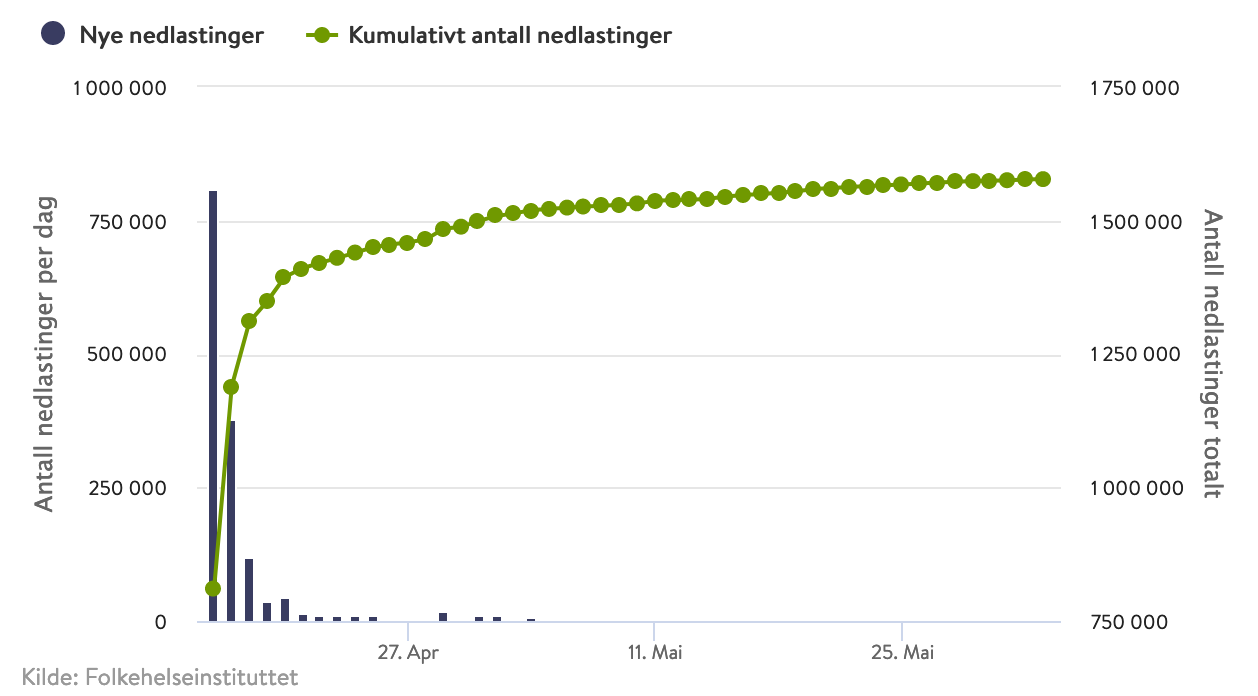}
		\caption{
			A graph showing the downloads of Smittestopp over time. Each bar represents one day of
      downloads. The dotted line represents total downloads. Taken from~\protect{\cite{ss-numbers}}.	
		}
	\end{center}
\end{figure}

In total, roughly 30\% of the Norwegian population downloaded the app, with the majority of downloads
happening the first few days after launch. In the following weeks, the number of users 
quickly flattened, and then started falling. By June, the number of active users had dropped to less than
14\% among Norwegians 16 years or older.\footnote{These numbers were calculated with user data from
NIPH~\cite{ss-numbers} and population data from Statistics Norway~\cite{norges-befolkning}. Note that the first number is an
estimate, due to the fact that re-downloads of the app by the same person cannot be precisely accounted for.}

It is fair to assume that the dwindling user base can be blamed on the large number of
highly critical articles published in Norwegian news media in the weeks following Smittestopp's launch,
\footnote{
  While such articles are too
  numerous to list, see~\cite{vg} for an early example.
}
together with the authorities'
inability or unwillingness to answer the worries of the populace.
The small user base, the complete lack of public trust in the application, 
and the low infection numbers in Norway at the time, were all factors leading to NDPA's temporary ban, as announced on
June 12 \cite{datatilsynet-varsel}. In total, less then ten warnings of close contact was issued by
the app to an end user~\cite{lessthanten}. Furthermore,
according to a source close to the project, none of the data collected ended up 
being processed for research before the ban was issued and the data deleted. 
While this may be comforting to privacy enthusiasts,
it also means that, by failing to gain the public's trust, the NIPH 
failed to achieve \emph{any} of their goals with the application. 

What could they have done differently?
First, and foremost, they could have chosen to pay attention to---and engage in---the
international discourse and developments on privacy preserving digital contact tracing that was happening at the time.

On March 19, a full month before the launch of Smittestopp, the European Data Protection Board (EDPB) 
released a statement on the processing of personal data in 
context of the pandemic \cite{edpb-statement}. The statement included the following:


\begin{quote}
	\emph{Personal data that is necessary to attain the objectives pursued should be
	processed for specified and explicit purposes.} [\dots]\
	\emph{The least intrusive solutions should always be preferred, taking into account the specific
	purpose to be achieved.}
\end{quote}

The previously mentioned Joint Statement on Contact Tracing was released on April 19, shortly after
Smittestopp's launch, and on April 21, EDPB published
guidelines for digital contact tracing, to be adopted throughout the EEA. 
They include~\cite[page 7]{edpb-guidelines}:

\begin{quote}
	\emph{In the context of a contact tracing application, careful consideration should be given to
	the principle of data minimisation and data protection by design and by default:}
	\begin{itemize}
		\item 	\emph{contact tracing apps do not require tracking the location of individual users. Instead, proximity
			data should be used;}
		\item 	\emph{as contact tracing applications can function without direct identification of individuals, appropriate
			measures should be put in place to prevent re-identification;}
		\item	\emph{the collected information should reside on the terminal equipment of the user and only the relevant
			information should be collected when absolutely necessary}.
	\end{itemize}
\end{quote}

Secondly, they could have realized from the start that, codes of ethics aside, nothing will be
achieved without the trust of the populace: without widespread adoption, digital
contact tracing is simply not effective. This was already realized by the authors of the 
Science-published article that originally suggested digital contact tracing as a tool to combat COVID-19~\cite[page 5]{ferretti2020quantifying}:

\begin{quote}
	\emph{Successful and appropriate use of the app relies on it commanding well-founded public trust
	and confidence. This applies to the use of the app itself and of the data gathered.}
\end{quote}

At this point, it should have been clear to the NIPH that development of a new version of
Smittestopp should commence immediately---a version that neither relied on nor collected GPS data, 
that was in line with international regulation, and that aimed to earn the public's trust. Instead, they seem to have ignored
the European legislation, and faced the criticism with an appeal to blind trust and
``dugnadsånd'',\footnote{A Norwegian term characterising a neighbourly and helpful spirit, used
repeatedly by then prime minister Erna Solberg at the outset of the pandemic.} with
the implication being that anyone who refused to download the app was being selfish at the expense of
those around them.

Then, on June 2, Amnesty International sent an open letter to Monica Mæland, then minister
of justice and public security, in the process sparking the biggest media storm so far.
The letter concludes~\cite{amnesty-letter}: 

\begin{quote}
	\emph{Amnesty International’s Security lab identified the following features of the app that are highly
	concerning from a privacy perspective and do not comply with several human rights standards
	outlined above:
	\begin{itemize}
		\item 	The app requires registration with a valid phone number. Thus, the operators of the app can
				tie any data upload to an identifiable individual.
		\item	The app collects GPS data. It stores a local copy, but also uploads this data to a central
				server. This allows operators of the app to track movement and location data of thousands of
				people who have the app installed. The Smittestopp app thus has the potential to be a mass
				surveillance tool.
		\item	The app also uploads all user data to third-party Microsoft servers, which appear to be
				operating in Ireland.
		\item	The app also performs Bluetooth-based contact tracing. Apps running on devices in the
				proximity will exchange the respective unique identifiers and store them locally, along
				with a timestamp and signal strength. These records are also uploaded to the central server.
				This data is thus neither anonymised, nor decentralised, allowing app operators to track
				users’ movements making it a privacy violation. Additionally, the use of unique identifiers
				could enable malicious actors to track users’ movements using a distributed network of
				Bluetooth sensors. This is a privacy risk.
	\end{itemize}
	Given the grave privacy risks to thousands of people, we wanted to alert you to this and urge you to
	immediately roll back the app in its current form and ensure that any contact tracing efforts
	are human rights respecting.}
\end{quote}

As we've seen, the app was discontinued less than two weeks later.

  \section{Conclusion}
  	In January 2018, The World Economic Forum published a Code of Ethics for Researchers~\cite{wef-coe}.
The document attempts to define a set of principles that may be adopted by all who do research, irrespective
of field of study. They are:

\begin{itemize}
	\item	\emph{Engage with the public.
	\item	Pursue the truth.
	\item	Minimize harm.
	\item	Engage with decision-makers.
	\item	Support diversity.
	\item	Be a mentor.
	\item	Be accountable.}
\end{itemize}
If we are to judge the parties based on these ``field-agnostic'' principles, it seems clear that the
NIPH failed on three points. 
First, they failed to engage with the public. Instead, they brushed aside complaints, and 
refused to accept criticism.

Second, in their ``pursuit of truth'', and in not realizing (or acknowledging) the 
importance and non-trivial nature of privacy preservation, they did not seek to minimize harm, even as the
harmful nature of their practice was pointed out to them.

Third, as we've seen, they failed to engage with decision-makers, completely ignoring legislation on
contact-tracing applications issude by the EU---legislation under which Norway, as a member of the EEU, was subject. 

To their credit, though, once the NDPA issued their
temporary ban on the processing of personal data in the app, the NIPH chose to immediately 
discontinue the app and delete all data gathered~\cite{ss-end}; as such, they stood accountable. 

On September 28, the Norwegian health minister announced that a new version of Smittestopp was in
development~\cite{ny-smittestopp}, 
this time relying on the Google and Apple Exposure Notification system, or
GAEN~\cite{gaen}.\footnote{
  In the time since this essay was written, the new application has been released, with very little
  fanfare and with no media storm in sight~\cite{ny-ss}. With the number of downloads 
  now surpassing one million, and with close to 5000 having used the app to notify that they have been
  infected, it's safe to say that the GAEN-based Smittestopp has been a relative success.~\cite{new-ss-numbers}
  (Further statistics, like the total number of users notified of close contact, are unavailable due
  to the privacy-preserving nature of the application.)
}
This architecture, 
based on one of the main proposals to come
out of the Pan-European Privacy-Preserving Proximity Tracing organization, is
decentralised, meaning that someone with access to the central servers will not learn anything about
its user base, and it is designed from the bottom up with the privacy of its users in
mind.\footnote{
  Whether it achieves a satisfying level of privacy remains however disputed. This goes both for the
  underlying protocol itself~\cite{vaudenay2020centralized}, and for GAEN in particular, as Google
  will upload data---including both personal and location data---on all Google Play Store users at 20
  minute intervals~\cite{simula-rapport}.
}

Due to severe restrictions on the Bluetooth capabilities of third-party applications on iPhone, it is
now generally agreed that for a decentralized contact-tracing app to be effective, it \emph{must} be based on
GAEN.\footnote{This, dubbed ``the Bluetooth problem'', was overcome in the original
Smittestopp design partly through the use of location data, and partly through the centralized 
design.}
Governments have pleaded with Apple to lift the restrictions on apps issued by official authorities 
for the purposes of contact tracing, but Apple won't budge, saying effectively that GAEN is their solution, 
and they can take it or leave it. To quote Michael Veale, one of the designers of the
architecture underpinning GAEN~\cite{veale},

\begin{quote}
	[GAEN is] \emph{great for individual privacy, but the kind of infrastructural power it enables should give us
	sleepless nights. Countries that expect to deal a mortal wound to tech giants by stopping them
	building data mountains are bulls charging at a red rag. In all the global crises, pandemics and
	social upheavals that may yet come, those in control of the computers, not those with the largest
	datasets, have the best visibility and the best---and perhaps the scariest---ability to change the
	world.} [\dots] \emph{Law should be puncturing and distributing this power, and giving it to individuals, communities and,
	with appropriate and improved human-rights protections, to governments.}
\end{quote}

In other words, pleading won't cut it.

  \section{Epilogue}
  	Then, just as one might think that the dust had settled for, that the axes were finally
buried and we were all ready to look ahead to a future of privacy-preserving digital contact tracing, Simula Research
Laboratories, the developer of Smittestopp,\footnote{And, through company ownership, the author's employer.}
decided to speak up.~\cite{kyrres-uttalelser}

\begin{figure}[h]
	\begin{center}
		\includegraphics[width=\textwidth]{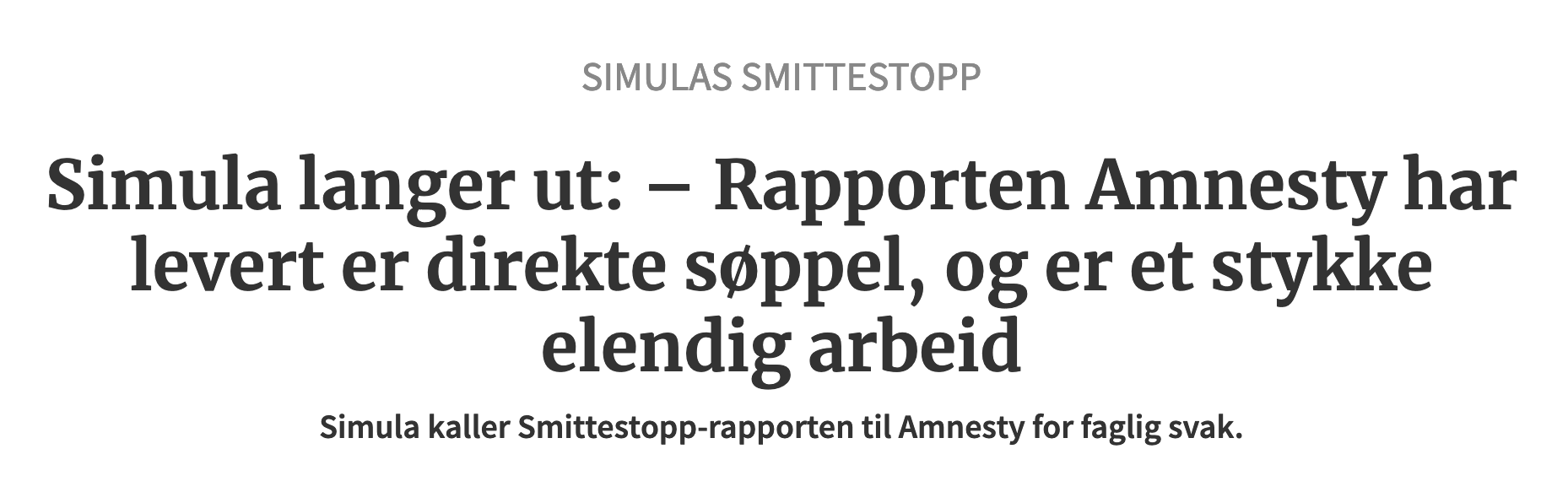}
		\caption{
      \label{fig:soeppel}
      The headline, taken from~\protect{\cite{kyrres-uttalelser}}, roughly translates to \emph{Simula lashes out: `Amnesty's
			report is nothing but trash, and a lousy piece of work.'}
		}
	\end{center}
\end{figure}

The statements of Fig.~\ref{fig:soeppel} are due to the corporate vice president of Simula, and were published on September
29, one day after the development of a new version was
announced.\footnote{Simula chose not to partake in the development of the new version of
Smittestopp.} He goes on to say that Amnesty has exploited the trust they hold with the public, either for the purpose of
bringing more attention to themselves, or to ``further an activist agenda''. Smittestopp, he claims,
satisfied almost all of Amnesty's principles, with the ``sole exception'' being that it was
gathering movement data on its users.

It is difficult to imagine what Simula was hoping to achieve with these statements. If
nothing else, they witness a severe disconnect from the international community that has grown
around constructing and analysing digital contact tracing. One would at the very
least have thought they were \emph{aware} of the many statements, guidelines, and legislative documents that
were published since March 2020, particularly given the ban issued by the NDPA, but the corporate
vice president's statements have cast even this into doubt. 
As a matter of fact, he seems to blame
Amnesty for NDPA's ruling, implying that the NDPA issued a ban on Smittestopp simply because it was the popular
thing to do at the time. 

If it is true that researchers have a duty to interact with decision-makers and the public,
then we can't help but wonder if there aren't better ways to do it.


  \ifeprintfull
    \clearpage
    \addcontentsline{toc}{section}{References}
  \fi

  \bibliographystyle{plain}
  \bibliography{references}

  \ifsubmission
    \clearpage
    \newgeometry{margin=1.0in}
    \appendix
  \else
    \appendix
  \fi

  \ifsubmission
    \clearpage
  \fi

\end{document}